# Ultrathin Stable Ohmic Contacts for High-Temperature Operation of $\beta$-Ga$_2$O$_3$ Devices

William A. Callahan[1,3], Edwin Supple[2], David Ginley[3], Michael Sanders[2], Brian P. Gorman[2], Ryan O'Hayre[2,a)], Andriy Zakutayev[3,b)],

[1] Advanced Energy Systems Graduate Program, Colorado School of Mines, Golden, Colorado, 80401, USA
[2] Department of Metallurgical and Materials Engineering, Colorado School of Mines, Golden, Colorado, 80401, USA
[3] National Renewable Energy Laboratory, Golden, Colorado 80401, USA

a) Electronic mail: rohayre@mines.edu
b) Electronic mail: andriy.zakutayev@nrel.gov

Beta gallium oxide (β-Ga$_2$O$_3$) shows significant promise in the high-temperature, high-power, and sensing electronics applications. However, long-term stable metallization layers for Ohmic contacts at high temperature present unique thermodynamic challenges. The current most common Ohmic contact design based on 20 nm of Ti has been repeatedly demonstrated to fail at even moderately elevated temperatures (300-400°C) due to a combination of non-stoichiometric Ti/Ga$_2$O$_3$ interfacial reactions and kinetically favored Ti diffusion processes. Here we demonstrate stable Ohmic contacts for Ga$_2$O$_3$ devices operating up to 500-600°C using ultrathin Ti layers with a self-limiting interfacial reaction. The ultrathin Ti layer in the 5nm Ti / 100nm Au contact stack is designed to fully oxidize while forming an Ohmic contact, thereby limiting both thermodynamic and kinetic instability. This novel contact design strategy results in an epitaxial conductive anatase titanium oxide interface layer that enables low-resistance Ohmic contacts that are stable both under long-term continuous operation (>500 hours) at 600°C in vacuum ($\leq 10^{-4}$ Torr), as well as after repeated thermal cycling (15 times) between room temperature and 550°C in flowing $N_2$. This stable Ohmic contact design will accelerate the development of high-temperature devices by enabling research focus to shift towards rectifying contacts and other interfacial layers.

# I. INTRODUCTION

$\beta$-$Ga_2O_3$ is a strong candidate for next-generation high-temperature electronic device manufacturing for both power and sensing applications. As a material it shows excellent figures of merit for high-voltage operation and high frequency switching, due to its large bandgap and high theoretical breakdown field [1–3]. Additionally, $\beta$-$Ga_2O_3$ is readily n-type dopable at shallow levels, with Si and Sn being the most common dopants[1]. From a manufacturing perspective, single-crystal $Ga_2O_3$ substrates can be grown by both Czochralski (CZ) and edge-defined film-fed growth (EFG), which allows for potential industrial scaling. Projective cost modeling suggests that as the technology develops, the economic viability and technological value of $Ga_2O_3$-based devices will continue to improve and be competitive with existing SiC and GaN technologies [4,5].

Reliable and stable Ohmic contacts are enabling for all types of $Ga_2O_3$-based devices. Currently, many groups use some variation of the commonplace Ti/Au metallization scheme, where most frequently a Ti layer of at least 20nm is applied, followed by a chemically protective and electrically conductive Au layer [6–18]. Variations that include additional diffusion barrier layers are also common [19–24]. This contact scheme is most frequently annealed at 470 °C for 90 seconds in nitrogen. Titanium is chosen as an interlayer due to its good adhesion to both semiconductors/oxides and to other metals, as well as its desirable electrical properties – Ohmic contacts of this design regularly demonstrate minimal degradation at or around room temperature operation. For $Ga_2O_3$-based devices, this contact scheme has demonstrated low specific contact resistances, on the order of $10^{-3} \Omega$ $cm^2$ for traditional n-type substrates [16,17], and between $10^{-5} - 10^{-6} \Omega$ $cm^2$ for surface treated substrates (e.g., ion implantation) [12,13].

While these contacts demonstrate acceptable stability at room temperature, few studies have examined the effects of long-term, high-temperature operation on their performance; to date, the few studies done reveal significant problems with rapid contact degradation even at relatively modest temperatures (300-400 °C). One group showed that a Si-implanted substrate treated with a reactive ion etch demonstrated stability with this metallization scheme over the course a 100+ hours thermal aging procedure at 300°C. Compared to this treated substrate, the contact resistance of an untreated metal-semiconductor junction increased by almost 40% over the same period [17]. Another group subjected a vertical Schottky device with a 20nm Ti / 100nm Au Ohmic contact to repeated thermal cycling, up to 410°C, and found that the series resistance of the device increased by several orders of magnitude [18]. This is problematic for applications where operation to 600°C is desired.

Degradation of this standard Ohmic contact design is hypothesized to be due, at least in part, to the formation of a 3-5 nm $Ti/TiO_x$ "defective" interfacial layer between the $Ga_2O_3$ and the Ti contact. This is supported by thermodynamic analysis, as annealing can provide conditions that are favorable for the formation of several different titanium oxides, resulting in redox reactions between the titanium and gallium oxide. This is thought to result in a gallium-rich sub-oxide layer in contact with a $TiO_x$ sub-oxide layer. Problematically, the remainder of the unreacted Ti either forms nanocrystals within the Au[16] or migrates through the Au to the outer surface over time, likely due to favorable thermodynamic driving factors (i.e., gradients in oxygen chemical potential), as well as facile Ti-diffusion kinetics that increase with temperature [18,25]. As such, electrical performance can be marred by the high Ti mobility and its redistribution during device operation. Both the formation of an oxide layer on top of the gold contact layer, as well as the formation of nanocrystalline scattering sites, are thought to reduce the performance and reliability of the Ohmic contact.

While Ti migration is thought to be detrimental to contact stability, the thin $Ti/TiO_x$ "defective" interfacial layer that forms between the $Ga_2O_3$ and the Ti contact can potentially be beneficial. Several studies of the oxidation state of Ti at the $Ti/Ga_2O_3$ interface suggest the presence of various $Ti_xO_{(2x-1)}$ Magnéli phases, which have been shown to be highly conductive and stable in oxidizing environments [26–29]. These findings suggest that leveraging and controlling the interfacial reaction while minimizing excess mobile Ti can potentially be used to improve contact stability.

In this report, we show that an ultra-thin titanium interlayer maximizes the completion of the reaction with the $Ga_2O_3$ substrate while minimizing subsequent Ti diffusion. We first compare the performance of a Ti/Au contact to (001) $Sn:Ga_2O_3$ with 5nm and 10nm of titanium through repeated thermal cycling (15 times) in a $N_2$ atmosphere. We find that the 10nm thick Ti contact shows both greater series resistance and inconsistent thermal and temporal behavior. In contrast, the 5nm thick Ti contact shows excellent stability and performance under long term temperature cycling. We then further validate the 5nm thick Ti contact design by fabricating a series of 12 double-Ohmic vertical devices which we subject to extended high-temperature thermal treatment (600°C) in a vacuum chamber ($\leq 10^{-4}$ Torr) for >500 hours. We find that all 12 devices show excellent performance, stability, and repeatability during this extended high-temperature exposure. Through TEM analysis of the thermally cycled samples, we find that high-temperature operation of the 5nm samples results in the formation of a highly crystalline, epitaxial titanium oxide interfacial layer which we hypothesize contributes to the reliable, conductive Ohmic contact behavior. Conversely, TEM investigation reveals incomplete oxidation of the thicker 10nm Ti-based contacts, which corelates with the unstable thermal and temporal behavior of these devices. This also likely explains the thermal instability of the even thicker 20nm Ti-based contact scheme commonly adopted in the field. We find that diffusion of Ti to the outer surface occurs in both 5nm and 10nm Ti samples, suggesting that its effects are secondary to those of the interfacial contact layer quality.

## II. EXPERIMENTAL METHODS

To test the Ohmic contacts, we employ both thermal cycling ("cycle") and long-term thermal holds at static temperature ("soaking") under an inert, flowing atmosphere ($N_2$) and vacuum ($\leq 10^{-4}$ Torr), respectively.

### A. *Sample Preparation*

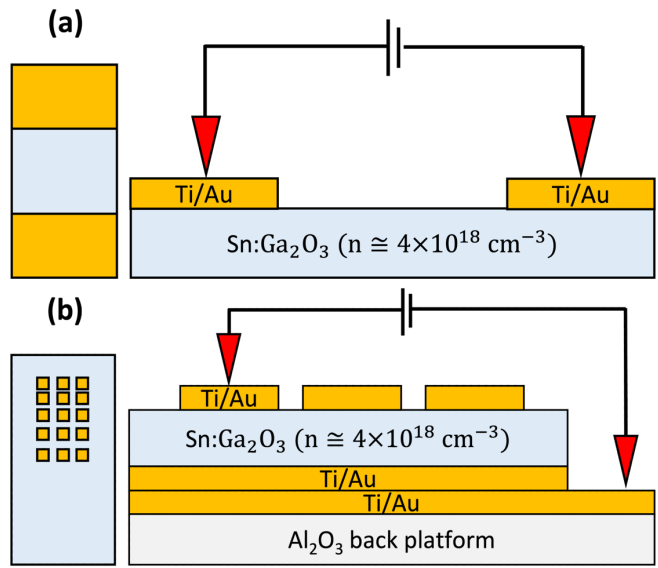

Figure 1: Top-down and profile views of device architecture and measurement scheme for (a) thermal cycling and (b) thermal soaking measurements. Contacts are comprised of a thin titanium layer (5 or 10nm for cycling; 5nm for soaking) capped by a 100nm Au layer. Carrier concentration determined from temperature-dependent Hall measurements.

Devices were fabricated using (001) Sn-doped $Ga_2O_3$ from Novel Crystal Technologies. Photoresist was removed from the as-delivered substrate via an organic wash, followed by a sulfuric acid/peroxide rinse. The bi-metal deposition of Ti/Au via e-beam was performed with a Temescal FC2000 Evaporation System in high-vacuum ($\leq$ $2 \times 10^{-6}$ Torr) without venting between layers. Contacts fabricated for thermal cycling consist of two large-area pads (average area of each $\cong 0.13$ cm$^2$) deposited on a 5mm × 10mm piece of Ga$_2$O$_3$ substrate in a lateral configuration. Contacts fabricated for thermal soaking consist of a monolithic large-area back contact, and a 3x5 matrix of 500μm square pads (area of each pad $= 2.5 \times 10^{-3}$ cm$^2$) in a linear transmission line method (LTLM) configuration as a front contact. Following contact deposition, samples were annealed via rapid thermal processing (ULVAC-RIKO MILA-3000 Rapid Thermal Processing Unit) at 550°C for 90 seconds in flowing N$_2$. On separate $10 \times 10$ mm substrates with metallized corners, the carrier concentration was determined via high-temperature Hall measurement from $50 - 400$ °C via a custom instrument [30]. Values were approximately constant in this temperature range at ~$4 \times 10^{18}$ cm$^{-3}$. See supplementary material at [URL will be inserted by AIP Publishing] for a more detailed depiction of these results, including resistivity and mobility.

## B.  *Electrical Characterization*

For thermal cycling, two-probe measurements (source I, measure V) were performed by a Keithley 236 SMU in an Instec HCP621G-PMH probe station under flowing nitrogen at 40 sccm. A single thermal cycle consisted of ramping from 25°C to 550°C and back down to 25°C in 75°C increments. Each temperature increment was allowed to equilibrate for 15 minutes before electrical measurements were performed.

Thermal soaking experiments were performed in a custom McAllister vacuum probe station with two probe arms, each equipped with 3-axis stepper motors. The stepper motor and temperature could be controlled externally, enabling sequential measurement of all devices without breaking vacuum. A vacuum level of $\leq 10^{-4}$ Torr was maintained by an Agilent TPS-compact vacuum pump. Electrical measurements were performed with the same Keithley 236 SMU. To initiate the long-duration thermal soak, the probe station was first ramped from room temperature to 600°C at approximately 50 °C per hour, with electrical measurements collected after each $75 - 90$°C increment.

Both probe stations utilized a silver stage with embedded thermocouples for a temperature feedback loop via the temperature control software. It was assumed that hold times (15 minutes for cycling, 1-2 hours for soaking) were long enough for thermal equilibration to occur between the stage and the device. Device design and testing configurations for both cycling and soaking are shown in Figure 1.

# III. RESULTS AND DISCUSSION

## A. Thermal Cycling

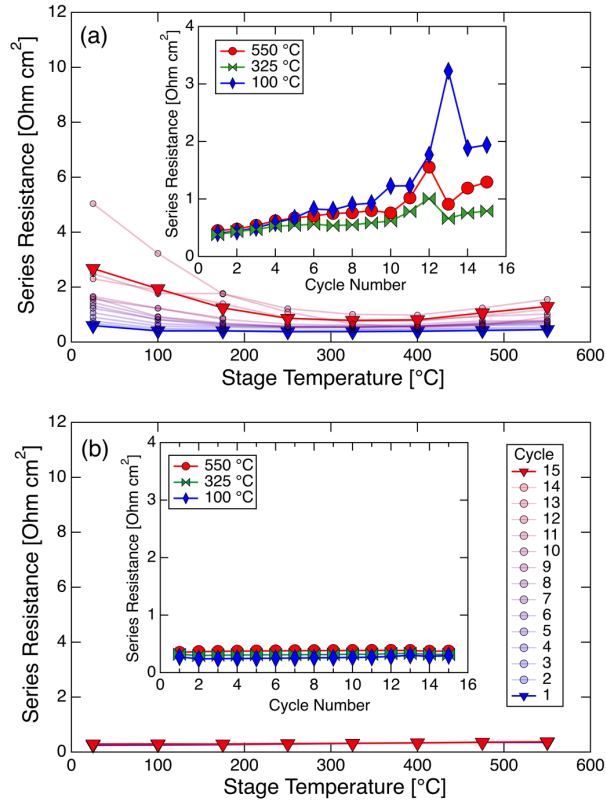

Figure 2: Thermal cycling experiments for (a) 10nm Ti and (b) 5nm Ti architectures, shown on the same scale bar and using the same legend. Insets show series resistance as a function of cycle number for 3 different temperatures.

Figure 2 summarizes the results of electrical measurements for 10nm and 5nm Ti contact architectures subject to extended thermal cycling. For both samples, the temperature was ramped from 25°C to 550°C and back down to 25°C in increments of 75°C. Electrical measurements were performed during each temperature increment after a 15-minute equilibration period, and a total of 15 cycles were performed. Total testing time was approximately 90 hours for each of the contact architectures.

Significant variability in the measured series resistance was observed for the 10nm sample, especially with increasing cycle number. The first 8 cycles produced relatively consistent resistance values that gradually increased from $\approx 0.6 \ \Omega \ cm^2$ (cycle 1) to $\approx 1.4 \ \Omega \ cm^2$ (cycle 8). Series resistance was highest at room temperature and trended towards a minimum at 325°C before gradually increasing again at still higher temperatures. After the first 8 cycles, the series resistance behavior became less consistent, particularly at lower temperatures, with values ranging between $\approx 1.5 - 5 \ \Omega \ cm^2$, representing an increase of 250-900% compared to the initial cycle behavior.

In contrast, the 5nm sample exhibited extremely stable and consistent behavior. Averaged over all cycles, the series resistance increased very slightly with increasing temperature (e.g. from $\approx 0.31 \pm 0.035 \ \Omega \ cm^2$ at room temperature to $\approx 0.32 \pm 0.032 \ \Omega \ cm^2$ at 550 °C), as well as with increasing cycle number. After 15 cycles, the room temperature series resistance increased from $\approx 0.26 \ \Omega \ cm^2$ (cycle 1) to $\approx 0.37 \ \Omega \ cm^2$ (cycle 15), a change of 44% or $0.11 \ \Omega \ cm^2$.

## B. Thermal Soaking

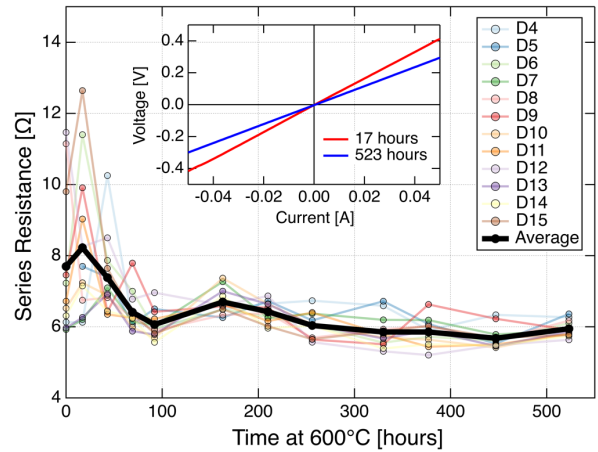

Figure 3): Time evolution of series resistance of 5nm Ti architecture for 12 devices on an absolute scale. Inset shows 12-device averaged IV curves at the start and end of the soak.

After reaching 600°C, the series resistance of the 12 devices were periodically measured every 1-3 days. An alumina platform with the same metallization scheme (5nm Ti / 100nm Au) was utilized to access the back Ohmic contact of the device. Shown in Figure 3, the average series resistance of the 12-devices decreased in magnitude from an initial value of $7.7 \pm 2.09$ Ω to a final value of $5.95 \pm 0.23$ Ω over the course of the >500-hour thermal soak at 600°C. Note that these values are reported in Ω instead of $\Omega \text{ cm}^{-2}$ as they include the probe tip resistance, leads, etc., which were found to contribute significantly to the overall resistance. Measurements of each of these contributions were attempted; however, they yielded results that proved too difficult to deconvolve. Hence, we report the lumped resistance values in their entirety. See supplementary material at [URL will be inserted by AIP Publishing] for full results of both TLM measurements and back platform resistances. The first ≈ 50 hours of thermal soaking produce the greatest variability; after this, all variations seen in the average series resistance across all devices are within one standard deviation of the mean. This behavior suggests a 'break-in' period occurs during the initial ~50hr period at temperature during which thermal or electrical processes drive the system towards a stable equilibrium.

## C. *TEM/EDS Analysis*

To understand this behavior, TEM lamellae were prepared from cycled samples of both the 5nm and 10nm Ti layer contacts using an FEI Helios Nanolab 600i, following standard lift-out techniques and using a 2kV clean final step to minimize sample damage[31]. The 5nm specimen was capped with carbon via permanent marker before FIB lift out to protect the top surface, while the 10nm specimen was capped only with electron-beam Pt GIS deposit followed by ion-beam Pt GIS deposit. The 5nm specimen was lifted out parallel to the substrate (010) plane while the 10nm specimen was lifted out parallel to (100). The TEM specimens were analyzed in an FEI Talos F200X Scanning Transmission Electron Microscope (STEM) at 200kV. A camera length of 98 mm was used for STEM imaging with spot size 9 for micrographs and spot size of 5 for Energy Dispersive Spectroscopy (EDS) mapping. Microprobe mode and spot size of 7 were used for diffraction mapping.

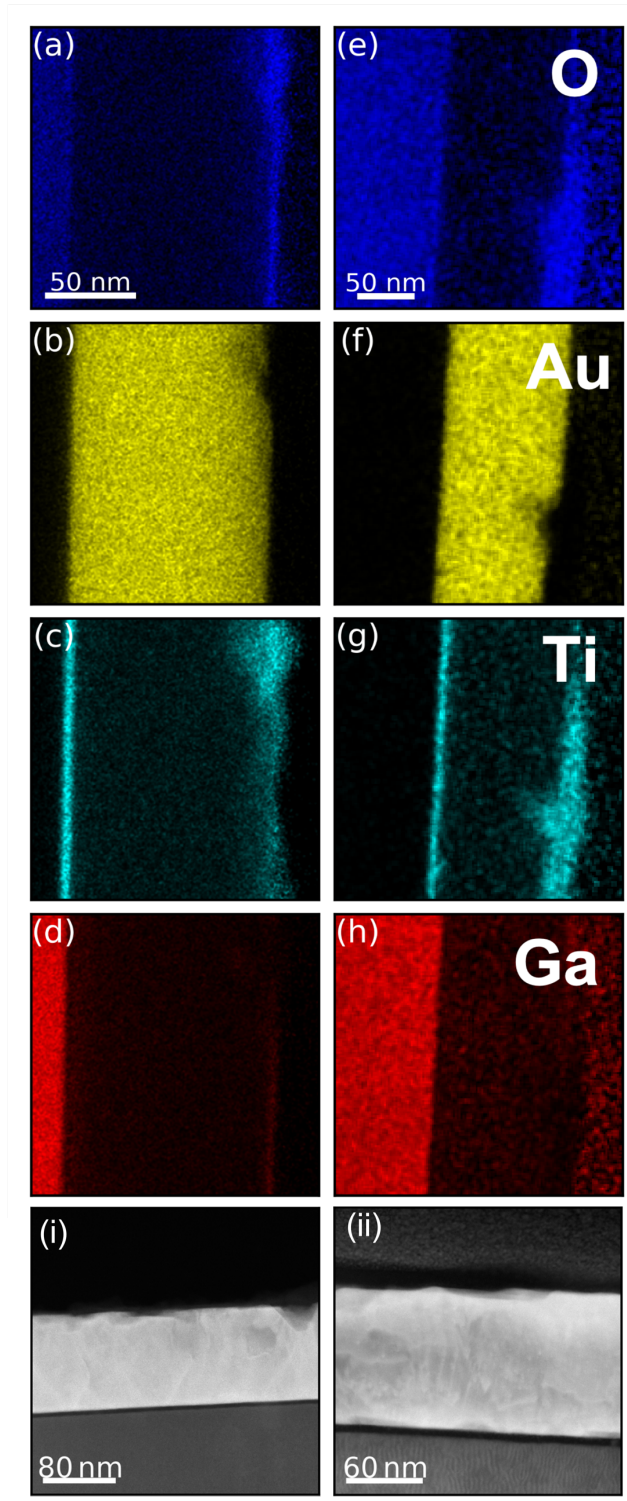

Figure 4): EDS map showing atomic concentration of a-d) the O, Au, Ti, and Ga content from the 5nm sample respectively, and e-h) the O, Au, Ti, and Ga content from the 10nm sample, respectively, with the substrate, Ti adhesion layer, Au layer, and capping material visible in the field of view. There are several similar features in both samples - Ga deposits on the Au in a thin even layer; and oxidized Ti migrates to the surface but in inconsistent clumps where there is a divot in the Au. Figures (i) and (ii) show STEM HAADF images of the 5nm Ti and 10nm Ti samples, respectively, by atomic number contrast. Note the difference in Ti layer thickness consistencies between samples.

Figure 4 shows EDS mapping of the two specimens reveals that in both samples, Ti and Ga have migrated through the gold layer to the top surface of the device. Gallium forms a very thin (< 5 nm), uniform layer across the top Au surface, while Ti is inconsistently distributed across the Au surface with some thicker regions and some regions where very little Ti is present. Intensity profiles across this wide field of view indicate that this outer Ti layer is oxidized. Since sample preparation was carried out using a Ga FIB, contamination giving the impression of Ga presence cannot be discounted. However, the 5nm Ti specimen provides good evidence that this Ga layer is real and not an artifact since the carbon capping contains no Ga and is sufficiently thick to protect the buried Au surface from any Ga implantation due to the milling process. Ga implantation from the side during sample thinning would be expected to be consistently dense throughout and mostly removed by the final 2kV polish.

The bottom two panels of Figure 4 show HAADF images of the 5nm and 10nm samples, contrasted by atomic number. The Ti contact appears as the thin dark layer in both TEM images. In both cases, the actual Ti layer thickness observed in the TEM cross sections is slightly smaller than nominal target thickness (~8nm for "10nm" contact and 4.5-4.7nm for "5nm" contact), demonstrating the need for a high degree of precision when fabricating contacts. Additionally, the Ti layer of the 10nm specimen shows slightly more thickness variation. The single-crystal $Ga_2O_3$ substrate underlying the 10nm Ti sample (bottom layer in the TEM image) has vertical striations most likely due to more aggressive specimen preparation curtaining the lamella or possibly due to differences in milling behavior of the differing crystallographic plane of the substrate.

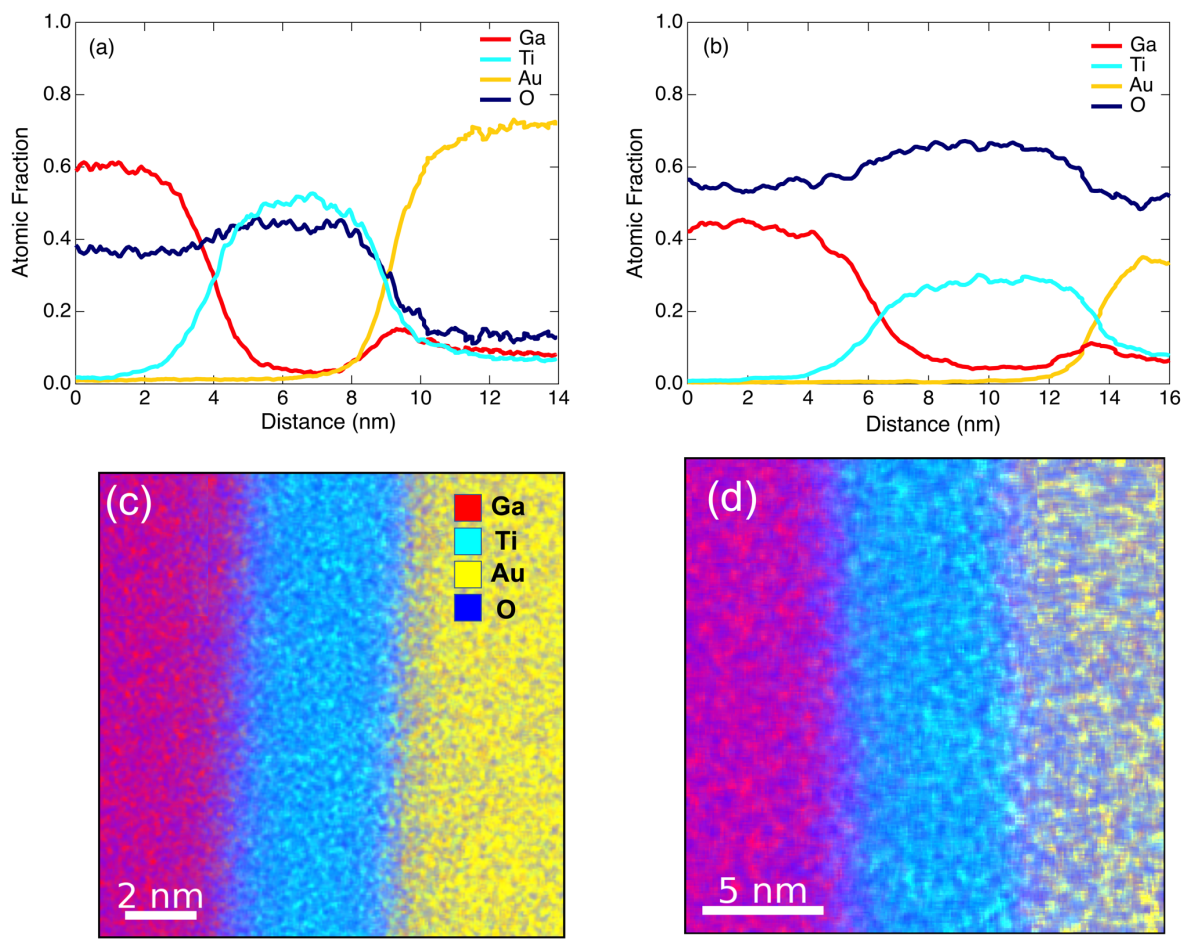

Figure 5 a) Atomic fraction EDS profile through the 5nm Ti layer, showing moderate oxygen concentration in the Ti layer and Ga concentration present at the Ti-Au interface. b) Atomic fraction EDS profile through the 10nm Ti layer, showing significant oxygen concentration that approaches $TiO_2$ stoichiometry, in the Ti layer and Ga present at the Ti-Au interface. c) EDS map showing distribution of Ga, Ti, Au, and O in 5nm sample across the Ti layer region. d) EDS map showing distribution of Ga, Ti, Au, and O in 10nm sample across the Ti layer region.

Figure 5 shows intensity profiles of the EDS signal from narrow field of view EDS maps at the Ti/ $Ga_2O_3$ interfacial layer. In both specimens, the Ti layer is oxidized, with the 10nm layer having a higher oxygen fraction than the 5nm layer. The oxygen fraction in the 10nm layer approaches the value expected for stoichiometric $TiO_2$, but the EDS map indicates significant oxygen content throughout the sample including in the gold layer; hence, conclusions related to the exact stoichiometry of the interfacial layer cannot be established with full certainty. The intensity profiles also show that there is Ga diffusion into the Ti layer and a small peak of higher Ga concentration between the Ti and the Au, further confirming that Ga is likely migrating through the samples, leading also to the Ga layer previously noted on the top Au surface of the samples.

## D. Diffraction analysis

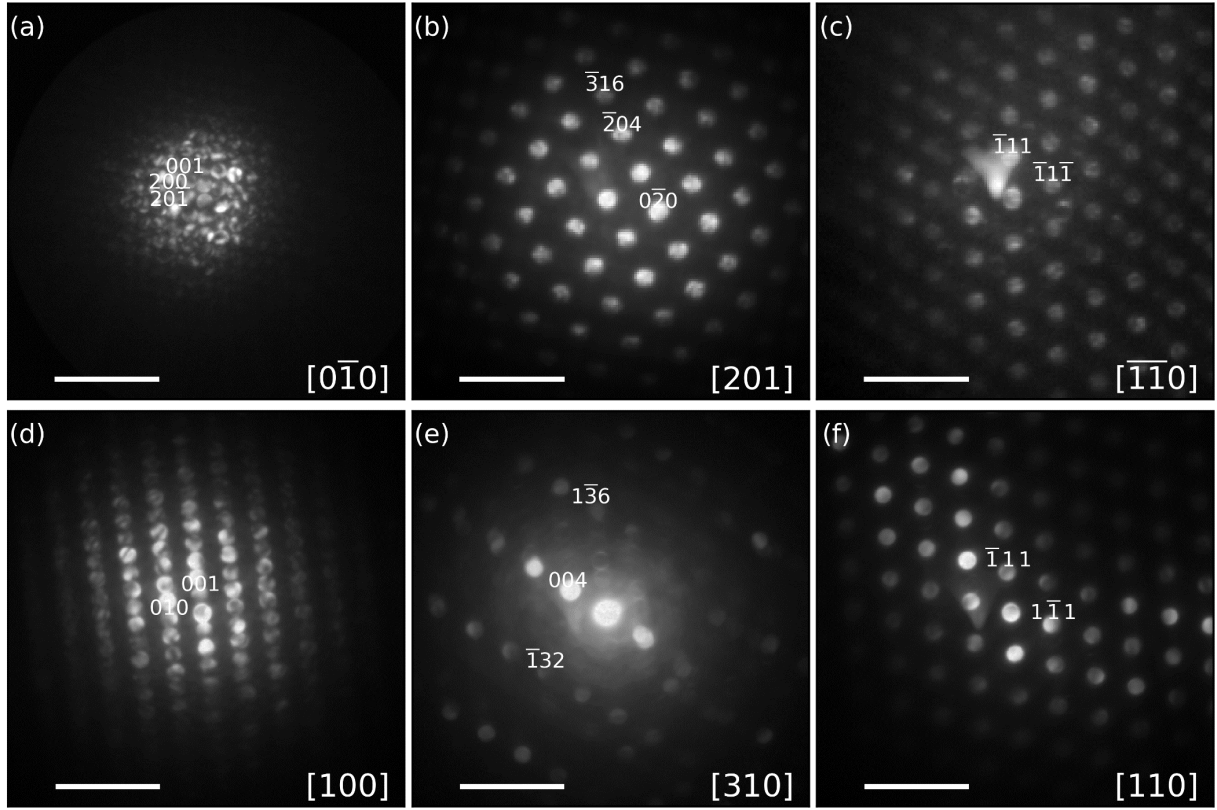

Figure 6): Diffraction patterns from representative areas of each sample, 5nm diffraction patterns in the top row and 10nm in the bottom row. Scale bar is 10 Å$^{-1}$ in all panels. a) Ga$_2$O$_3$ viewed on [0$\bar{1}$0] zone axis. This diffraction pattern was recorded separately from the rest of the map to improve signal-to-noise and contrast around the closely spaced diffraction spots. b) Ti layer identified as anatase TiO$_2$ phase viewed on [201] zone axis with ($\bar{3}$16) co-oriented with the Ga$_2$O$_3$ (001) reflection. c) Au [110] diffraction pattern with ($\bar{1}$11) reflection co-oriented with the Ga$_2$O$_3$ (001) reflection. d) Ga$_2$O$_3$ viewed on [100] zone axis. e) Ti layer identified as anatase TiO$_2$ phase viewed on [310] zone axis with (1$\bar{3}$6) co-oriented with Ga$_2$O$_3$ (001) reflection. Note the significant amount of amorphous character and secondary phases present in this diffraction pattern compared to the other patterns within this sample and the other sample. f) Au [110] diffraction pattern with ($\bar{1}$11) co-oriented with Ga$_2$O$_3$ (001) reflection.

Diffraction pattern maps were taken of each sample aligned with the zone axes of the planes parallel to the lamellae, i.e. [100] and [0$\bar{1}$0] for the 10nm and 5nm samples, respectively. Representative diffraction patterns from each layer of each sample are shown in Figure 6. The (001) substrate reflection is co-oriented with the surface in both cases. The titanium layer diffraction patterns were indexed against elemental Ti as well as rutile, anatase, and brookite $TiO_2$ phases. Best matches were the [310] and [201] anatase zones for the 10nm and 5nm samples, respectively. The anatase diffraction patterns of both samples indicate {$\bar{3}16$} is co-oriented with the growth plane. The 10nm sample shows substantially more amorphous character (diffuse intensity rings around the direct beam) and diffraction spots from a secondary phase in the diffraction patterns of the Ti layer. In contrast, the substrate and the Au diffraction patterns lack any signs of amorphous character in either sample. The Au diffraction patterns reveal textured polycrystalline structure. ⟨110⟩ zones make up most of the specimen with (111) generally oriented in the growth direction. The thin Ga layer indicated by EDS between the Ti and Au was not evident in these diffraction maps.

Separate diffraction maps were taken from the $TiO_x$ clusters on the top Au surface of the 5nm sample only. This material is polycrystalline with grains on the scale of 1nm. The grains are not epitaxial and only some diffraction patterns are on a discernable zone. Both rutile and brookite diffraction patterns are identifiable, but with significant amorphous character. However, this amorphous contribution could come largely from the protective capping added during sample preparation rather than the sample itself. Although not investigated, we presume that the $TiO_x$ clusters formed on the top Au surface of the 10nm sample are similarly structured with a mix of rutile, brookite, and amorphous $TiO_x$.

The diffraction mapping results indicate epitaxial growth of anatase in both cases, but the 10nm layer has some amorphous character and secondary phase and does not complete its transformation from elemental Ti to fully oxidized anatase TiO$_2$. The implied epitaxial relationship $(001)_{Ga_2O_3}$ || $(\bar{3}16)_{TiO_2}$ and $[0\bar{1}0]_{Ga_2O_3}$ || $[201]_{TiO_2}$ is at first improbable. Firstly, the anatase [201] and [310] zones are not perpendicular; they are 78 degrees separated. The anatase diffraction patterns off-zone from [310] do however appear very similar to [310] even at large mistilts. Simulations of diffraction patterns 90 degrees perpendicular to $[210]_{anatase}$ (i.e. 12 degrees from [310]) nevertheless share many of the same reflections as [310] and include the $(\bar{3}16)$ reflection. So, the 10nm Ti diffraction pattern is only close to [310] and not fully on zone.

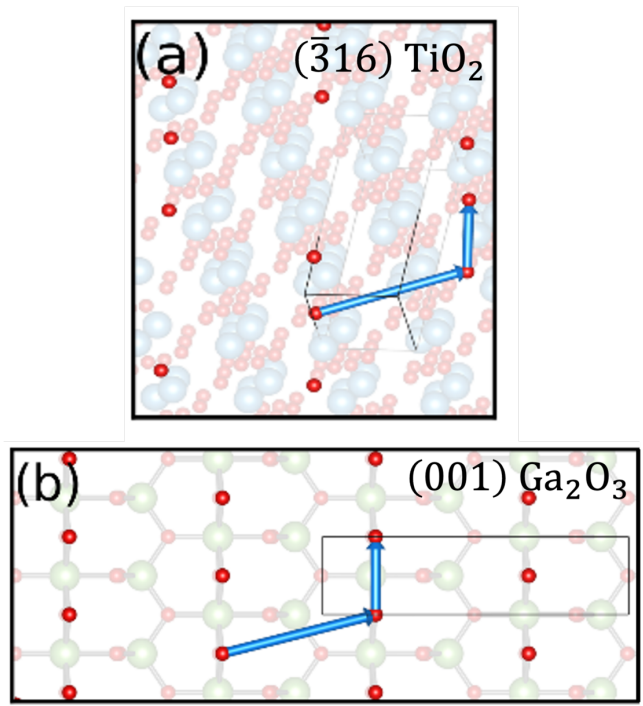

Figure 7: Cross sections of a) anatase TiO$_2$ $(\bar{3}16)$ and b) Ga$_2$O$_3$ (001) planes showing oxygen sublattice in bright red and semi-transparent underlying crystal structure. Corresponding geometry for possibility of epitaxial growth is marked with blue arrows. The crystals are oriented in the proposed epitaxial relationship.

We also consider the oxygen-containing $(\bar{3}16)_{anatase}$ planes relative to the $Ga_2O_3$ substrate, shown in Figure 7. Both planes have a similar offset dumbbell motif; the O-O spacing in the [210] direction of anatase $TiO_2$ as well as the [010] direction of $Ga_2O_3$ is approximately 3.1Å. The O-O spacing between adjacent dumbbells is also similar at 6.2Å and 6.4Å for $TiO_2$ and $Ga_2O_3$, respectively. There is a reasonable atomistic matching pattern to explain this epitaxial growth pattern and diffraction evidence supporting this co-orientation in both specimens and viewed in separate directions.

To summarize, TEM analysis shows that in both the 5nm and 10nm cases, the Ti layer forms epitaxial anatase phase $TiO_2$; Ti migrates to the surface and forms unevenly distributed titanium oxide patches; and Ga also migrates to both the Ti-Au interface and the surface and forms a thin, evenly distributed layer. The key difference between the two specimens is that the 5nm anatase layer is fully crystalline whereas the 10nm anatase layer contains a substantial amount of amorphous $TiO_x$ second phase that affects the performance of the Ohmic contact.

## IV. SUMMARY AND CONCLUSIONS

We have successfully fabricated Ohmic contacts to Sn:Ga$_2$O$_3$ using an ultra-thin layer of Ti (5nm) with a Au capping layer (100nm). These contacts show remarkable stability and excellent Ohmic performance after both extensive thermal cycling between 25 - 550°C in N$_2$ and long-term thermal soaking at 600°C for ≥ 500 hours under vacuum conditions. Through TEM analysis, we show that the 5nm Ti layer is sufficiently thin that it completely transforms to an epitaxial, highly conductive anatase titanium oxide layer which provides a stable Ohmic contact. In contrast the 10nm Ti layer does not fully react, leading to a partially amorphous TiO$_x$ layer that does not enable a stable Ohmic contact. While Ti is found on the outer surface of the Au layer in both 5nm and 10nm Ti samples, the homogeneity of the interlayer at the Ga$_2$O$_3$ interface appears to have a greater impact in the overall performance and stability of the contact. Integration of this ultrathin Ohmic contact design will aid development of high-temperature devices, by specifically shifting research focus to stable Schottky metals, p-type heterojunction materials, and other functional layers.

# ACKNOWLEDGMENTS

This work was authored by the National Renewable Energy Laboratory (NREL), operated by Alliance for Sustainable Energy, LLC, for the U.S. Department of Energy (DOE) under Contract No. DE-AC36-08GO28308. Funding provided by the Office of Energy Efficiency and Renewable Energy (EERE) Advanced Manufacturing Office. This work was supported in part by National Science Foundation (NSF) Award No. 2125899. The views expressed in the article do not necessarily represent the views of the DOE or the U.S. Government.

# AUTHOR DECLARATIONS

**Conflicts of Interest**

The authors have no conflicts to disclose

**Author Contributions**

<u>Ohmic contact design and testing</u>:

W. Callahan – Conceptualization (equal), Data Curation (lead), Formal Analysis (lead), Investigation (lead), Methodology (equal), Software (lead), Visualization (lead), Writing/Original Draft Preparation (lead), Writing/Review & Editing (equal)

D. Ginley – Conceptualization (equal), Funding Acquisition (equal), Methodology (equal), Writing/Review & Editing (equal)

M. Sanders – Conceptualization (equal), Methodology (equal)

R. O'Hayre – Conceptualization (equal), Funding Acquisition (equal), Methodology (equal), Writing/Review & Editing (equal)

A.Zakutayev – Conceptualization (equal), Funding Acquisition (equal) , Methodology (equal), Project Administration, (lead), Writing/Review & Editing (equal)

<u>TEM analysis:</u>

E. Supple – Data Curation (lead), Formal Analysis (lead), Methodology (equal), Visualization (lead), Writing/Review & Editing (equal)

B. Gorman – Methodology (equal), Writing/Review & Editing (equal)

## DATA AVAILABILITY

The data that supports the findings of this study are available within the article [and its supplementary material].